\documentclass[twocolumn,english,pra,showpacs,superscriptaddress,longbibliography]{revtex4-1}
\usepackage{amsfonts}
\usepackage{amsmath}
\usepackage{graphics}
\usepackage{amssymb}
\usepackage{physics}
\usepackage[colorlinks, citecolor=blue,linkcolor=red]{hyperref}
\usepackage{color}
\usepackage{hyperref}
\usepackage{textcomp}
\usepackage[rightcaption]{sidecap}
\usepackage{subfigure}
\usepackage{float,csquotes,mathtools}
\usepackage[rightcaption]{sidecap}
\usepackage{graphicx}
\usepackage{lipsum}

\begin{document}
\title{Nonreciprocal Control of the Goos--H\"{a}nchen Shift via the Barnett Effect in Cavity Magnomechanics}
\author{Shah Fahad}
\affiliation{Department of Physics, Zhejiang Normal University, Jinhua, Zhejiang 321004, China}
\author{Gao Xianlong}
\email {gaoxl@zjnu.edu.cn}\affiliation{Department of Physics, Zhejiang Normal University, Jinhua, Zhejiang 321004, China}

\setlength{\parskip}{0pt}
\setlength{\belowcaptionskip}{-10pt}
\begin{abstract} 
We propose a theoretical scheme for realizing a tunable nonreciprocal Goos–H\"{a}nchen shift (GHS) in a hybrid cavity magnomechanical system. The setup consists of a rotating yttrium iron garnet sphere embedded in a microwave cavity, with magnetic-dipole and magnetostrictive interactions mediating magnon–photon and magnon–phonon couplings, respectively. Owing to the Barnett effect, the magnon frequency acquires a rotation-induced shift whose sign can be reversed by changing the direction of the bias magnetic field. We show that the output probe spectrum exhibits a Fano resonance, while the associated GHS responds asymmetrically to opposite field directions, providing a controllable mechanism for nonreciprocal beam shifts. The magnon–photon and magnon–phonon interactions are found to affect the GHS in opposite ways, while the cavity length offers an additional degree of tunability. These results provide a route toward magnetically reconfigurable microwave photonic devices and sensitive detection of Barnett-induced effective fields.
\end{abstract}
\date{\today}
\maketitle
\section{Introduction}
Nonreciprocity---the asymmetric response of a system to opposite propagation directions---enables directional control of light transmission. This property is of significant interest in complex quantum networks and quantum communication~\cite{Shoji2014, Flamini2018}. The Sagnac effect in spinning resonators has been widely employed to explore nonreciprocal quantum phenomena, including quantum steering and entanglement~\cite{JiaoPRL2020, GuanPRA2024, Jiao2022}, squeezing~\cite{Guo2023}, and photon blockade~\cite{Huang-PRL}, with substantial experimental progress also reported~\cite{Maayani2018}. In addition, magnon Kerr nonlinearity can induce nonreciprocity through reversal of the bias magnetic-field direction~\cite{Wang2016Kerr}, enabling nonreciprocal bipartite and tripartite entanglement~\cite{Chen2023NE}. Nonreciprocal microwave-field transmission has also been demonstrated in quantum magnomechanical~\cite{Ullah2024} and cavity magnonic systems through Kerr nonlinearity~\cite{Kong2019}. More recently, the Barnett effect has been proposed as a mechanism to realize nonreciprocal magnon blockade~\cite{Huang2024Nonreciprocal}, opening new opportunities to investigate nonreciprocal quantum phenomena in macroscopic systems. 

The Barnett effect arises from rotation-induced alignment of magnetic moments and has been observed in ferromagnetic insulators~\cite{Kani2022, Barnett1915, Ono2015, Bretzel2009} and nuclear spin systems~\cite{Arabgol2019, Chudo2014}. It has been utilized to study rotational vacuum friction~\cite{Manjavacas2010}, enable remote magnetization control~\cite{Davies2024Phononic}, and detect the angular-momentum compensation point~\cite{Imai2019}. Building on these developments, the Barnett effect has recently been shown to enable the generation of bipartite and tripartite entanglement in cavity magnomechanical (CMM) systems~\cite{Lu2025}. Motivated by these findings, we investigate the influence of the Barnett effect on the Goos--H\"{a}nchen shift (GHS) in a CMM system, providing a route to its tunable control through mechanical rotation.

The CMM system integrates microwave photon, magnon, and phonon modes within a unified hybrid platform, bridging quantum information, magnonics, quantum optics, and cavity quantum electrodynamics (QED)~\cite{Zhang2015CavityQED, zuo_2024, Lachance2019, Li2020HybridMagnonics, Soykal2010}. The coherent interactions among these modes give rise to a variety of phenomena, including cavity-magnon polaritons~\cite{Cao_2015_Exchange, Yao_2015_Theory}, magnon bistability~\cite{Wang2016Kerr, Hyde_2018_Direct}, Bell states~\cite{Yuan2020}, quantum chaos~\cite{Peng2024}, squeezed states and entanglement~\cite{YU_2020, Qiu_2022, Li_2019_sqeez, Li_2022}, mechanical bistability~\cite{Shen2022}, and controllable magnonic switching~\cite{Hao_2024_controllabe}. They further enable rich interference phenomena in the microwave response, including magnon-induced transparency (MIT), absorption (MIA), and magnomechanically induced transparency (MMIT)~\cite{Kamran_2020, Li_2020_phase-control, Munir_2023, Zhang_2016_cavity}, analogous to optomechanically induced transparency in cavity optomechanical systems~\cite{Hou_2015, Stefan2010, Agarwal2010}. Beyond these interference phenomena, the phase response of CMM systems provides a means to control beam shifts, particularly the GHS. 

The GHS describes the lateral displacement of a reflected beam along the interface, arising from the angular dependence of the reflection phase~\cite{Goos-1947, Wang2008, Zia_2010_Coherent, Zia_2015_PRA, Shui_2019}. Owing to its sensitivity to the reflection phase, the GHS has found broad applications in various fields, including humidity sensing~\cite{Wang2016Humidity}. temperature sensing~\cite{Lu_2022, Chen2007}, Surface characterization~\cite{Le-PRL}, neutron optics~\cite{Haan-PRL, McKay-2025}, acoustics~\cite{Declercq2004}, waveguide theory~\cite{White1977}, and seismology~\cite{WANG2015}. Nonreciprocal GHSs have recently attracted considerable interest as a promising means of achieving directional beam-displacement control. Various platforms, including graphene-coated gyroelectric slabs~\cite{XU2016}, coupled magnetic chains~\cite{Ma2022}, Weyl semimetals~\cite{Yuliang2026}, parity--time-antisymmetric atomic lattices~\cite{Liu2025}, magnetic plasmonic gradient metasurfaces~\cite{Wu2019}, and cavity optomagnonic systems~\cite{Deng2026}, have been explored to realize nonreciprocal GHSs. However, the reported optomagnonic realization of nonreciprocal GHS has focused on an optical configuration based on photon--magnon interactions~\cite{Deng2026}. The GHS has also been investigated in CMM systems~\cite{waseem_Goos_2024, MUNIR2026}; nevertheless, the potential of the Barnett effect, which provides a controllable rotation-induced frequency shift, for manipulating the GHS and enabling nonreciprocal control remains unexplored.

In this paper, we investigate the manipulation of the GHS of the reflected probe field through the Barnett shift in a CMM system using the transfer-matrix method in conjunction with the stationary-phase approach. When the YIG sphere rotates under a bias magnetic field, the Barnett effect induces a rotation-dependent frequency shift of the magnon mode. Reversing the sign of the Barnett-induced frequency shift leads to distinct probe-field absorption and reflection-phase responses, enabling nonreciprocal control of both the sign and magnitude of the GHS without changing the incident conditions. In the absence of the Barnett shift, the GHS exhibits a relatively small magnitude, whereas introducing the Barnett shift significantly enhances the GHS and produces opposite-signed responses for opposite shift directions. We further show that the magnon--photon and magnon--phonon couplings exhibit distinct and opposite effects on the GHS in the absence and presence of the Barnett shift. In addition, the intracavity thickness provides an additional degree of freedom for controlling the nonreciprocal GHS. These results demonstrate the potential of Barnett-shift-mediated GHS control for tunable nonreciprocal beam manipulation and magnetically reconfigurable microwave photonic applications.

We organize the remainder of this paper as follows. Section II establishes the theoretical model, derives the effective optical susceptibility, and obtains the GHS using the stationary-phase approximation. Section III is devoted to the main results, with particular emphasis on the effects of the Barnett-induced frequency shift on the output probe-field spectra and GHS. Finally, Sec. IV provides a brief conclusion.
\section{Theoretical Model}
\subsection{System Hamiltonian and Heisenberg-Langevin equations}
\begin{figure}[tp]
\includegraphics[width=0.97\linewidth]{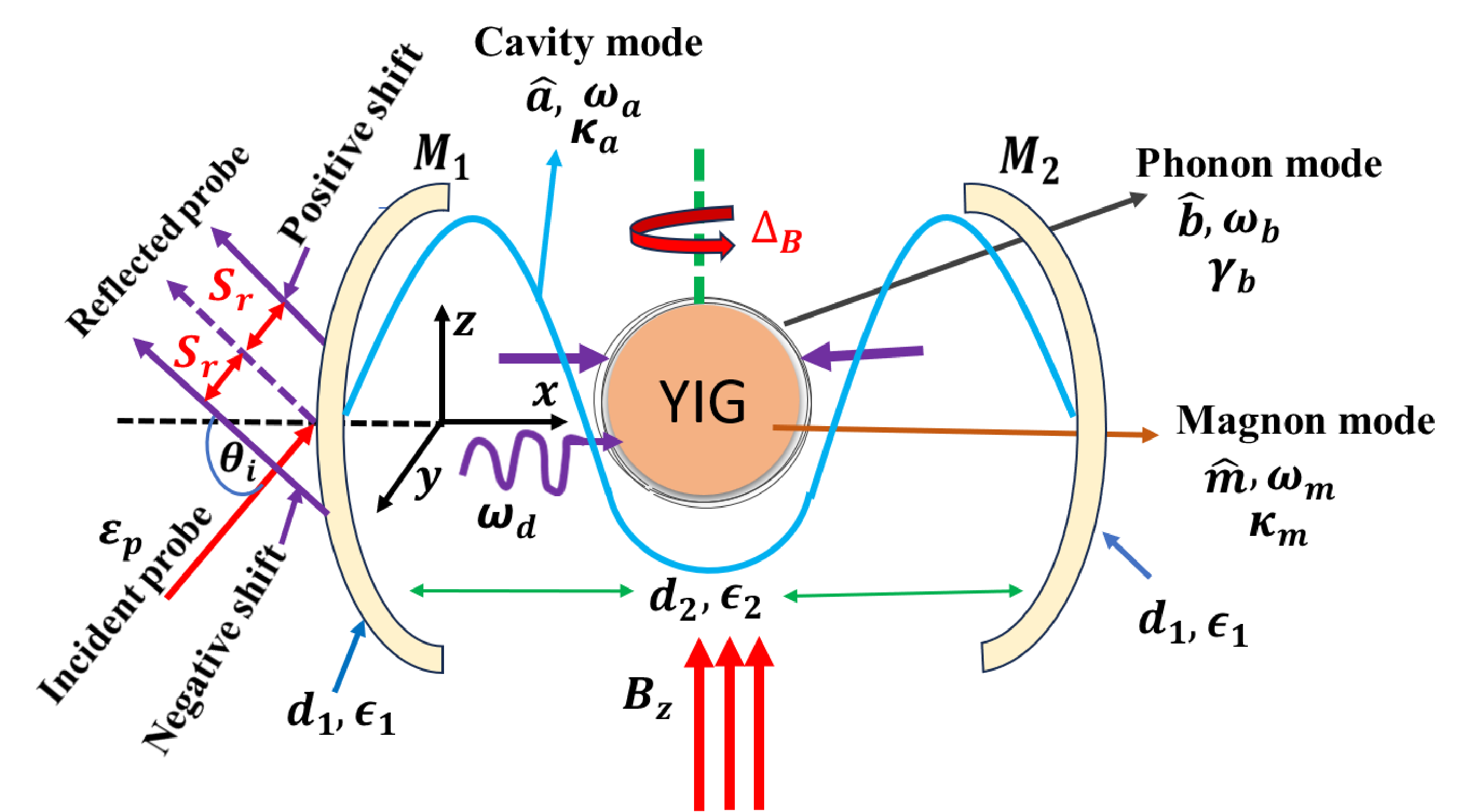}
\caption{Schematic of the YIG-based CMM system. The system comprises a cavity photon mode $\hat{a}$, a magnon mode $\hat{m}$, and a phonon mode $\hat{b}$, characterized by frequencies $(\omega_a,\omega_m,\omega_b)$ and decay rates $(\kappa_a,\kappa_m,\gamma_b)$, respectively. The magnon frequency is controlled by the static bias field $B_z$, while rotation of the YIG sphere at angular frequency $\Delta_B$ induces a Barnett field $H_B$ and consequently shifts the magnon resonance. The photon--magnon interaction is mediated by magnetic-dipole coupling, whereas the magnon--phonon interaction arises from magnetostriction and is enhanced by a microwave drive of frequency $\omega_d$ applied along the $x$-axis. The bias ($B_z$), drive ($B_x$), and cavity ($B_y$) magnetic fields are mutually perpendicular. A probe field $\mathcal{E}_p$ impinges on the partially reflecting mirror $M_1$ at an incidence angle $\theta_i$, and the resulting reflected field exhibits a positive or negative GHS $S_r$. The perfectly reflective mirror $M_2$ is placed at a distance $d_2$ from $M_1$.}
\label{fig1}
\end{figure}
We consider a CMM system comprising a microwave cavity photon mode ($\hat{a}$, resonance frequency $\omega_a$, decay rate $\kappa_a$), in which a YIG sphere is embedded (Fig.~\ref{fig1}). The YIG sphere simultaneously supports a uniform magnon mode ($\hat{m}$, resonance frequency $\omega_m$, decay rate $\kappa_m$) and a phonon mode ($\hat{b}$, resonance frequency $\omega_b$, decay rate $\gamma_b$). The coupling between cavity photons and magnons arises from the magnetic-dipole interaction, whereas the magnon--phonon coupling is mediated by the magnetostrictive interaction~\cite{Zhang_2016_cavity}. A uniform bias magnetic field $B_z$ is applied along the $z$-axis to tune the magnon resonance frequency, while the YIG sphere is assumed to rotate about the same axis with angular frequency $\Delta_B$. Owing to the Barnett effect~\cite{Bretzel2009, Ono2015, Barnett1915, Kani2022, Lu2025}, this rotation induces an effective Barnett field $B_B=\Delta_B/\gamma$ (gyromagnetic ratio $\gamma$), thereby modifying the magnon resonance frequency. Consequently, the Barnett effect shifts the magnon resonance frequency from $\omega_m$ to $\omega_m\pm\Delta_B$, reflecting the conservation of angular momentum. For a fixed counterclockwise rotation of the YIG sphere, the sign of the Barnett-induced frequency shift can be reversed by changing the direction of the applied magnetic field from $+z$ to $-z$, corresponding to $\Delta_B>0$ and $\Delta_B<0$, respectively. The system comprises two nonmagnetic mirrors, $M_1$ and $M_2$, separated by an intracavity length $d_2$. Mirror $M_1$ partially reflects, while $M_2$ completely reflects. Both mirrors possess the same thickness $d_1$ and permittivity $\epsilon_1$, while the intracavity medium has an effective permittivity $\epsilon_2$.

The system Hamiltonian in a frame rotating at the driving frequency $\omega_d$ is given by ($\hbar=1$)~\cite{Lu2025}:
\begin{equation}
\begin{aligned}
\hat{H} &= \Delta_a \hat{a}^{\dagger}\hat{a} + (\Delta_m + \Delta_B)\hat{m}^{\dagger}\hat{m} + \omega_{b}\hat{b}^{\dagger}\hat{b}\\
&+ g_{ma}(\hat{a}^{\dagger}\hat{m} + \hat{a}\hat{m}^{\dagger}) 
+ g_{mb} \hat{m}^\dagger\hat{m}(\hat{b}^{\dagger} + \hat{b}) \\ 
&+i\Omega_{d}(\hat{m}^{\dagger} - \hat{m}) +i\mathcal{E}_{p} \left( \hat{a}^\dagger e^{-i\delta_p t} - \hat{a} e^{i\delta_p t} \right),
\end{aligned}
\label{Main-H}
\end{equation}
where $\Delta_{a,m} = \omega_{a,m}-\omega_{d}$ and $\delta_{p}=\omega_{p}-\omega_{d}$ are the cavity (magnon)–drive and probe–drive detunings, respectively. Here $\hat{a}$ ($\hat{a}^\dagger$), $\hat{m}$ ($\hat{m}^\dagger$) and  $\hat{b}$ ($\hat{b}^\dagger$) are the annihilation (creation) operators of the cavity, magnon, and phonon modes, respectively, while $g_{ma}$ ($g_{mb}$) represents the magnon--photon (magnon--phonon) coupling. A microwave drive with amplitude $\Omega_d = \sqrt{5N}\gamma B_0/4$~\cite{Li2018} (magnetic field amplitude $B_0$ and total number of spins $N$) enhances the magnon-phonon (magnonmechanical) interaction. Meanwhile, a weak probe field with amplitude $\mathcal{E}_p = \sqrt{2P_p\kappa_a/\hbar\omega_p}$ (power $P_p$, decay rate $\kappa_{a}$, and frequency $\omega_p$) drives the cavity mode.

The system dynamics are governed by the Heisenberg–Langevin equation
\begin{equation}
\dot{\hat{\mathcal{C}}} = i [\hat{H}, \hat{\mathcal{C}}]-\Gamma\hat{\mathcal{C}}+ \mathcal{N}
\end{equation}
where $\hat{\mathcal{C}} \in \{\hat{a},\hat{m},\hat{b}\}$, $\Gamma$ denotes the corresponding decay rate, and $\mathcal{N}$ represents the input vacuum and Brownian noise terms. We obtain:
\begin{equation}
\begin{aligned}
\dot{\hat{a}} &= - (i \Delta_a + \kappa_a) \hat{a} - i g_{ma} \hat{m} + \mathcal{E}_p e^{-i \delta_p t} +\sqrt{2\kappa_{a}}\,\hat{a}_{\mathrm{in}},  \\
\dot{\hat{m}} &= - [i (\Delta_m + \Delta_B) + \kappa_m] \hat{m} - i g_{ma} \hat{a} - i g_{mb}\hat{m} (\hat{b}^\dagger +\hat{b})\\& + \Omega_d+\sqrt{2\kappa_{m}}\,\hat{m}_{\mathrm{in}}, \\
\dot{\hat{b}} &= - (i \omega_b + \gamma_b) \hat{b} - i g_{mb} \hat{m}^\dagger \hat{m} + \hat{\zeta},
\end{aligned}
\label{HLEqs}
\end{equation}
where $\kappa_a$, $\kappa_m$, and $\gamma_b$ are the cavity photon, magnon, and phonon decay rates, respectively, and $\hat{a}_{\mathrm{in}}$, $\hat{m}_{\mathrm{in}}$, and $\hat{\zeta}$ represent the corresponding noise operators. The noise operators have zero mean and obey the correlation functions~\cite{Lu2025, Xiong2015}: 
\begin{equation}
\begin{aligned}
\langle \hat{a}_{\mathrm{in}}^{\dagger}(t)
\hat{a}_{\mathrm{in}}(t') \rangle=\langle\hat{m}_{\mathrm{in}}^{\dagger}(t)\hat{m}_{\mathrm{in}}(t') \rangle=0,\\
\langle \hat{a}_{\mathrm{in}}(t) \hat{a}_{\mathrm{in}}^{\dagger}(t') \rangle=\langle \hat{m}_{\mathrm{in}}(t)
\hat{m}_{\mathrm{in}}^{\dagger}(t') \rangle=\delta(t-t'),\\
\langle \hat{\zeta}(t)\hat{\zeta}(t') \rangle=\frac{\gamma_b}{\omega_b}\int \frac{d\omega}{2\pi}e^{-i\omega(t-t')}
\left[\coth\left(\frac{\hbar\omega}{2k_B T}\right)+1\right],
\end{aligned}
\end{equation}
where $T$ denotes the temperature of the thermal bath and $k_B$ represents the Boltzmann constant. For a high mechanical quality factor ($Q_b=\omega_b/\gamma_b\gg1$),  $\hat{\zeta}(t)$ becomes delta-correlated~\cite{Vitali-PRL}: $\frac{1}{2}\left\langle
\hat{\zeta}(t)\hat{\zeta}(t') + \hat{\zeta}(t')\hat{\zeta}(t)
\right\rangle=\gamma_b(2N_b+1)\delta(t-t')$, where $N_b=[\exp(\hbar\omega_b/k_BT)-1]^{-1}$ is the mean thermal phonon number. Taking the expectation values of the operators $C(t) \equiv \langle \hat{C}(t) \rangle$ ($C = a, m, b$)~\cite{Xiong2015}, Eq.~(\ref{HLEqs}) reduces to the semiclassical equations of motion:
\begin{equation}
\begin{aligned}
\dot{a} &= - (i \Delta_a + \kappa_a) a - i g_{ma} m + \mathcal{E}_p e^{-i \delta_p t}  \\
\dot{m} &= - [i (\Delta_m + \Delta_B) + \kappa_m] m - i g_{ma} a - i g_{mb} m (b^\ast + b)\\& + \Omega_d, \\
\dot{b} &= - (i \omega_b + \gamma_b) b - i g_{mb} m^\ast m.
\end{aligned}
\label{Simplified form}
\end{equation}
Under the weak-probe approximation, we write $\mathcal{C}=\mathcal{C}_{s}+\delta\mathcal{C}$, where $\mathcal{C}\in\{a,m,b\}$, with $\mathcal{C}_{s}$ and $\delta\mathcal{C}$ denoting the steady-state components and first-order fluctuations, respectively. The steady-state solutions follow as:
\begin{equation}
\begin{aligned}
a_s =  \frac{-i g_{ma}m_s}{i \Delta_a + \kappa_a}, \\
m_s = \frac{-ig_{ma}a_{s} + \Omega_d}{[i (\tilde{\Delta}_m + \Delta_B) + \kappa_m ]},\\
b_s =  \frac{-i g_{mb}|m_s|^2}{i \omega_b + \gamma_b},
\end{aligned}
\end{equation}
where $\tilde{\Delta}_m=\Delta_m+g_{mb}(b_s+b_s^\ast)$ represents the effective detuning of the magnon mode. Retaining only the linear terms in the fluctuations, the resulting equations take the form:
\begin{equation}
\begin{aligned}
\delta\dot{a} &= - (i \Delta_a + \kappa_a) \delta a - i g_{ma} \delta m + E_p e^{-i \delta_p t}, \\
\delta\dot{m} &= - [i (\tilde{\Delta}_m+ \Delta_B) + \kappa_m] \delta m - i g_{ma} \delta a - i g_{mb} m_s\delta b, \\
\delta\dot{b} &= - (i \omega_b + \gamma_b) \delta b - i g_{mb} m_s^\ast \delta m.
\end{aligned}
\label{Ist-order-fluctuations-1}
\end{equation} 
Within the rotating-wave approximation, the rapidly oscillating counter-rotating terms $\delta b^\dagger$ \text{and} $\delta m^\dagger$ are neglected.
\subsection{Effective optical susceptibility}
To obtain the effective susceptibility of the intracavity medium, we transform the linear fluctuation amplitudes into a slowly varying frame via $\delta q \rightarrow \delta q e^{-i\nu_q t}$, where $(q,\nu_q)\in \left\{(a,\Delta_a),(m,\tilde{\Delta}_m),(b,\omega_b)\right\}$. Under the red-sideband condition $\Delta_a = \tilde\Delta_m = \omega_b$~\cite{Agarwal2010}, the linearized equations reduce to
\begin{equation}
\begin{aligned}
\delta\dot{a} &= -\kappa_a \delta a - i g_{ma} \delta m + \mathcal{E}_p e^{-i x t}, \\
\delta\dot{m} &= - (i\Delta_B + \kappa_m) \delta m - i g_{ma} \delta a - iG_{mb}\delta b, \\
\delta\dot{b} &= -\gamma_b \delta b - iG_{mb}^\ast \delta m.
\end{aligned}
\label{Simplified form}
\end{equation}
where $x=\delta_p-\omega_b$ is the effective detuning in the rotating frame, and $G_{mb}=g_{mb}m_s$ is the effective magnomechanical coupling strength. We employ the standard harmonic ansatz for the linearized fluctuations $\delta{\mathcal{C}}=\delta \mathcal{C}_1 e^{-ix t} + \delta \mathcal{C}_2 e^{ix t} \ ( \mathcal{C}=a,m,b)$ from which the first-order sideband amplitude $\delta a_1$ of the CMM system under a weak probe field follows as
\begin{equation}
\delta a_1=\frac{\mathcal{E}_{P}}{(\kappa_a -ix)+ \frac{g_{ma}^2({\gamma_b -ix})}{(\kappa_m +i\Delta_{B} -ix)({\gamma_b -ix}) + |G_{mb}|^2}}
\end{equation}
The $\delta a_2$ component corresponds to the counter-rotating (four-wave-mixing) sideband at frequency $\omega_p-2\omega_d$ and is therefore neglected in the probe-frequency response. The effective susceptibility $\chi$ is then defined through the output probe field $E_{T}$ as~\cite{waseem_Goos_2024, Fahad2026, Deng2026}
\begin{equation}
\chi \equiv E_T=\frac{\kappa_a \delta a_{1}}{\mathcal{E}_{p}}.\label{optical-sus}
\end{equation}
Here, $\chi=\chi_r+i\chi_i$ is a complex response function. Its real ($\chi_r$) and imaginary ($\chi_i$) parts characterize the absorptive and dispersive responses of the cavity to the probe field, respectively~\cite{waseem_Goos_2024, MUNIR2026, Li_2016Transparency, Chen_2023}. The effective permittivity $\epsilon_2$ describes the optical response of the intracavity medium and is related to the susceptibility through $\epsilon_2=1+\chi$.
\subsection{Goos–H\"{a}nchen shift}
We investigate the GHS $S_r$ associated with the reflection of the probe field from $M_1$. This shift originates from the angular dependence of the reflection phase and can be determined using the stationary-phase approximation. Within this framework, the probe field is assumed to be well collimated with a narrow angular spectrum ($\Delta K \ll K$). The GHS experienced by the reflected probe beam is therefore given by~\cite{Artmann-1948, Li-2003PRL}
\begin{equation}
S_{r} = -\frac{\lambda_{p}}{2\pi} \frac{d\phi_{r}}{d\theta_{i}}.
\label{phase}
\end{equation}
Here, $\lambda_p$ corresponds to the probe-field wavelength, while $\phi_r$ represents the phase of the reflection coefficient $R(k_z,\omega_p)$. $k_z=(2\pi/\lambda_p)\sin\theta_i$ is the wavenumber along the $z$ direction, and $\theta_i$ is the angle of incidence. Explicitly, the GHS can be expressed as~\cite{Wang2005}
\begin{equation}
\begin{split}
S_{r} = -\frac{\lambda_p}{2\pi} \frac{1}{|R(k_{z},\omega_{p})|^2}
\left\{\operatorname{Re}[R(k_{z},\omega_{p})] \frac{d \operatorname{Im}[R(k_{z}, \omega_{p})]}{d\theta_{i}}\right. \\
\left. - \operatorname{Im}[R(k_{z}, \omega_{p})] \frac{d \operatorname{Re}[R(k_{z}, \omega_{p})]}{d\theta_{i}}
\right\}.
\label{GHS} 
\end{split}
\end{equation}
To evaluate $R(k_z,\omega_p)$ appearing in Eq.~(\ref{GHS}), we employ the transfer-matrix formalism~\cite{Wang2008}:
\begin{equation}
R(k_{z}, \omega_{p}) =
\frac{q_{0}(T_{22}-T_{11})-\left(q_{0}^{2}T_{12}-T_{21}\right)}
{q_{0}(T_{22}+T_{11})-\left(q_{0}^{2}T_{12}+T_{21}\right)},
\label{Reflection-coefficent}
\end{equation}
where $q_{0}=\sqrt{\epsilon_{0}-\sin^{2}\theta_{i}}$, and $T_{ij}$ ($i,j=1,2$) represent the corresponding entries of the $2\times 2$ transfer matrix $T(k_z,\omega_p)$. The total transfer matrix for the three-layer structure is expressed as~\cite{Wang2008}
\begin{widetext}
\begin{equation}   
T(k_{z}, \omega_{p}) = \mathbb{M}_{1}(k_{z}, \omega_{p}, d_{1}) \mathbb{M}_{2}(k_{z}, \omega_{p}, d_{2})\mathbb{M}_{1}(k_{z}, \omega_{p}, d_{1}) =\begin{pmatrix}
T_{11} & T_{12} \\
T_{21} & T_{22}
\end{pmatrix},
\end{equation}
\end{widetext}
where $\mathbb{M}_j(k_z,\omega_p,d_j)$ is the transfer matrix of the $j$th layer, given by
\begin{equation}
\mathbb{M}_{j}(k_{z},\omega_{p},d_{j}) =
\begin{pmatrix}
\cos[k_{x}^{j} d_j] & i \sin[k_{x}^{j} d_{j}] k/ k_{x}^{j} \\
i\sin[k_{x}^{j} d_{j}]k_{x}^{j}/k & \cos[k_{x}^{j} d_{j}]
\end{pmatrix}, 
\end{equation}
where $k=\omega_p/c$ is the vacuum wavenumber, with $c$ denoting the speed of light, and $k_x^j=k\sqrt{\epsilon_j-\sin^2\theta_i}$ corresponds to the $x$-directed component of the probe-field wavenumber in the $j$th layer. Here, $d_j$ and $\epsilon_j$ denote the thickness and permittivity of the $j$th layer ($j=1,2$), respectively.
\begin{figure}
\centering
\includegraphics[width=0.95\linewidth]{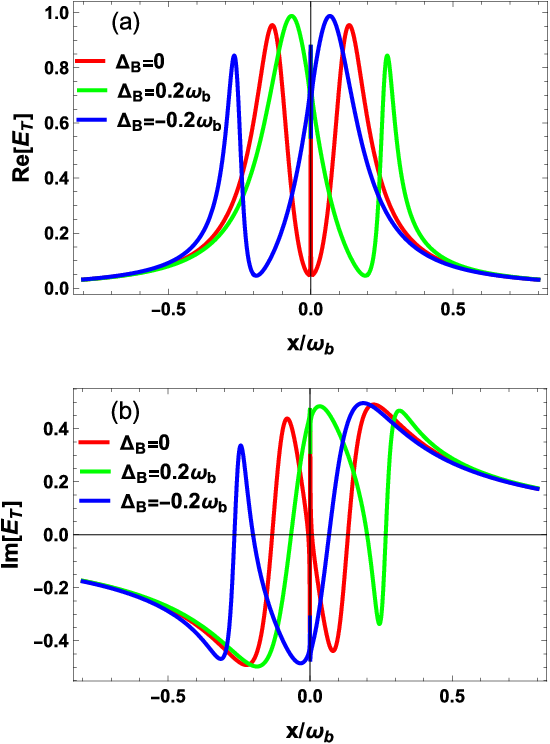}
\caption{(a) Absorption spectra ($\mathrm{Re}[E_T]$) and (b) dispersion spectra ($\mathrm{Im}[E_T]$) of the probe field as functions of the normalized effective detuning $x/\omega_b$. The curves correspond to different values of the Barnett frequency shift: $\Delta_B = 0$ (red), $0.2\omega_b$ (green), and $-0.2\omega_b $ (blue). Fixed parameters: $\gamma_b/2\pi=150~\mathrm{Hz}$, $\kappa_a/2\pi=2.1~\mathrm{MHz}$, $g_{ma}/2\pi=2~\mathrm{MHz}$, $G_{mb}/2\pi=0.1~\mathrm{MHz}$, $\kappa_m/2\pi=0.1~\mathrm{MHz}$, and $\omega_b/2\pi=15~\mathrm{MHz}$.}
\label{fig2}
\end{figure}
\begin{figure*}
\centering
\includegraphics[width=0.95\linewidth]{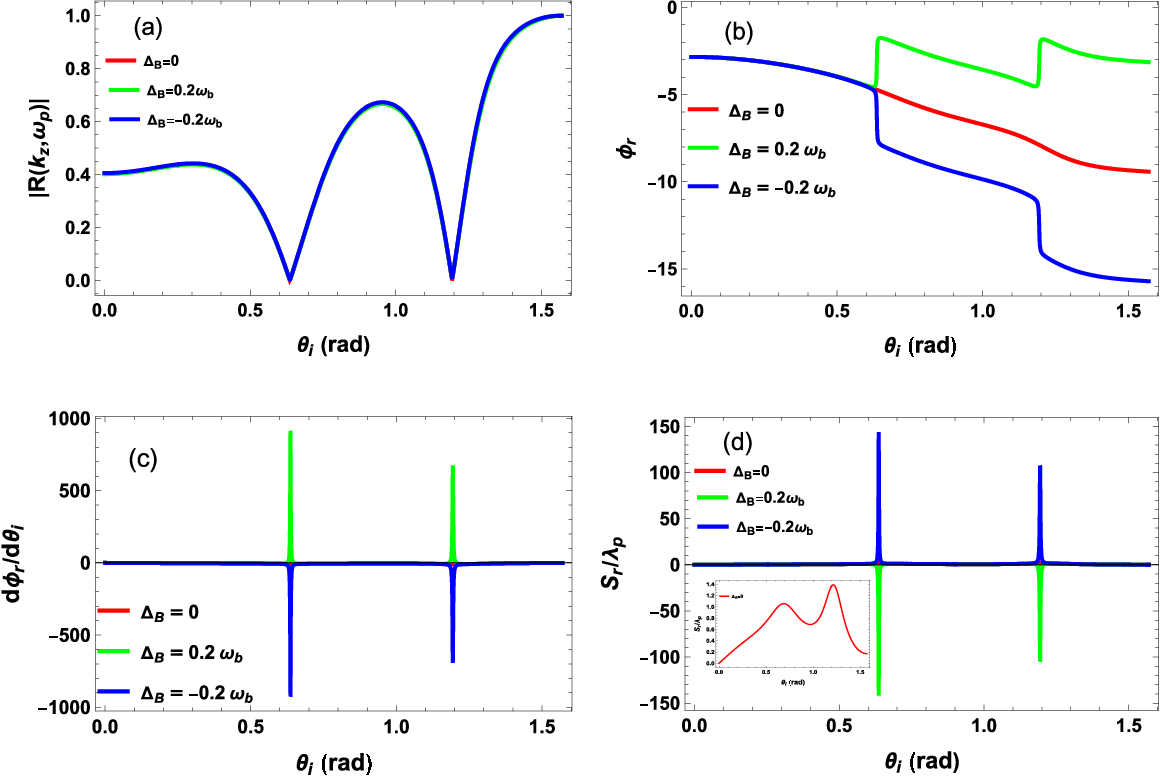}
\caption{(a) Magnitude of the reflection coefficient $|R(k_z,\omega_p)|$, (b) unwrapped phase of the reflection coefficient $\phi_r$, (c) phase derivative $d\phi_r/d\theta_i$, and (d) normalized GHS $S_r/\lambda_p$ as functions of the incident angle of the probe field $\theta_i$ at resonance ($x=0$). The curves correspond to Barnett frequency shifts of $\Delta_B=0$ (red), $0.2\omega_b$ (green), and $-0.2\omega_b$ (blue). The inset in Fig~\ref{fig3} (d) shows the $S_{r}/\lambda_p$ dependence on $\theta_{i}$ for $\Delta_B = 0$ (red) at $x=0$.  Fixed parameters: $\gamma_b/2\pi=150~\mathrm{Hz}$, $\kappa_a/2\pi=2.1~\mathrm{MHz}$, $\kappa_m/2\pi=0.1~\mathrm{MHz}$, $\omega_b/2\pi=15.0~\mathrm{MHz}$, $g_{ma}/2\pi=2~\mathrm{MHz}$, $G_{mb}/2\pi=0.1~\mathrm{MHz}$, $d_{1} = 4.0~\mathrm{mm}$, $d_{2} = 45~\mathrm{mm}$, $\epsilon_{0}=1.0$, $\epsilon_{1}= 2.2$, and $\omega_p/2\pi=13.2~\mathrm{GHz}$.}
\label{fig3}
\end{figure*}
\section{Results and discussion}
In this section, we perform the numerical analysis using experimentally accessible parameters~\cite{Harder_PRL, Zhang_2016_cavity}: $\omega_p/2\pi=13.2~\mathrm{GHz}$,  $g_{ma}/2\pi=2~\mathrm{MHz}$, $\omega_b/2\pi=15~\mathrm{MHz}$, $\kappa_m/2\pi=0.1~\mathrm{MHz}$, $\kappa_a/2\pi=2.1~\mathrm{MHz}$, $\gamma_b/2\pi=150~\mathrm{Hz}$, and $G_{mb}/2\pi=0.1~\mathrm{MHz}$. The YIG sphere has diameter $D=250~\mu\mathrm{m}$, $\gamma/2\pi=28~\mathrm{GHz/T}$, and spin density $\rho=4.22\times10^{27}~\mathrm{m}^{-3}$. To ensure dynamical stability, the driving field is restricted to $B_0\leq0.5~\mathrm{mT}$, corresponding to $G_{mb}/2\pi\leq1.5~\mathrm{MHz}$~\cite{Lu_2021_Ep}. For the GHS calculation, we set $\epsilon_0=1$, $\epsilon_1=2.2$, $d_1=4~\mathrm{mm}$, and $d_2=45~\mathrm{mm}$~\cite{Li_2020_phase-control, Zhang_2016_cavity}.

We start our numerical analysis by investigating the effect of the Barnett frequency shift $\Delta_B$ on the absorptive ($\mathrm{Re}[E_T]$) and dispersive ($\mathrm{Im}[E_T]$) responses of the output probe field, as described by Eq.~(\ref{optical-sus}). Figure~\ref{fig2}(a) shows the $\mathrm{Re}[E_T]$ as a function of the normalized effective detuning $x/\omega_b$. For a fixed rotation direction of the YIG sphere, a bias magnetic field applied along $+z$ ($-z$) induces a positive (negative) $\Delta_B$ via the Barnett effect~\cite{Bretzel2009, Ono2015, Barnett1915, Kani2022, Lu2025}. In the absence of the Barnett shift ($\Delta_B=0$, red curve), the spectrum exhibits a symmetric MMIT response centered at zero detuning, arising from the hybridization induced by the magnon--photon and magnon--phonon couplings. Introducing $\Delta_B$ shifts the spectral response in a sign-dependent direction: a positive shift ($\Delta_B=+0.2\omega_b$, green curve) displaces the transparency feature toward positive $x/\omega_b$, while a negative shift ($\Delta_B=-0.2\omega_b$, blue curve) moves it toward negative $x/\omega_b$. Reversing the sign of $\Delta_B$ thus reverses the direction of the spectral displacement, demonstrating direction-sensitive control of the MMIT response via the Barnett frequency shift. These modified resonance profiles originate from the altered interference among the coupled photon, magnon, and phonon modes, yielding a Fano-like spectral response~\cite{Kamran_2020}. A corresponding displacement of the dispersive response is also observed in $\mathrm{Im}[E_T]$ [Fig.~\ref{fig2}(b)], confirming that the Barnett shift provides simultaneous control over both the absorptive and dispersive probe-field responses.
\begin{figure}
\centering
\includegraphics[width=0.95\linewidth]{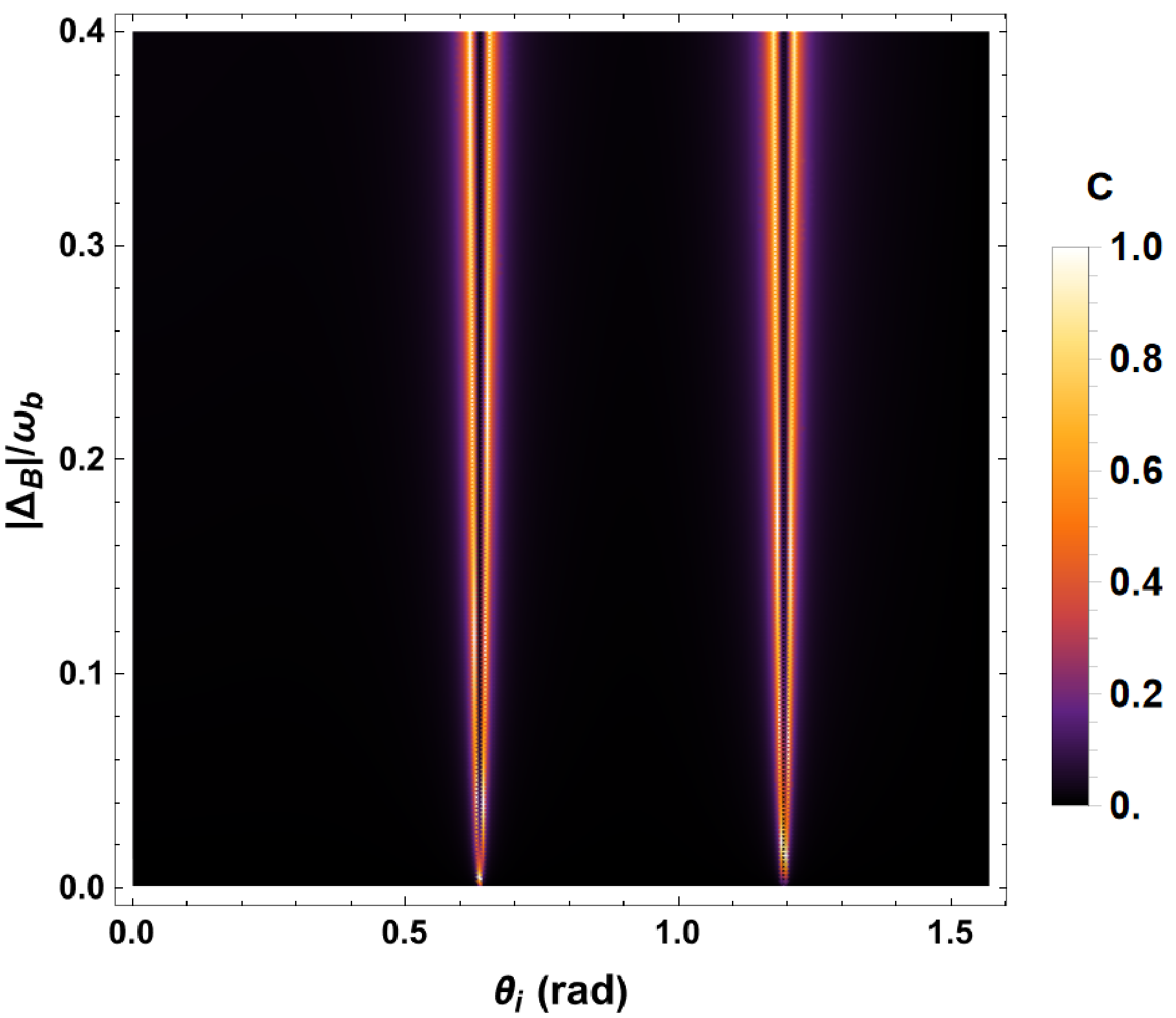}
\caption{Contrast ratio $C$ versus the probe-field incident angle $\theta_i$ at resonance ($x=0$). Other parameters are as in Fig.~\ref{fig3}.}
\label{fig4}
\end{figure}
\begin{figure*}
\centering
\includegraphics[width=0.98\linewidth]{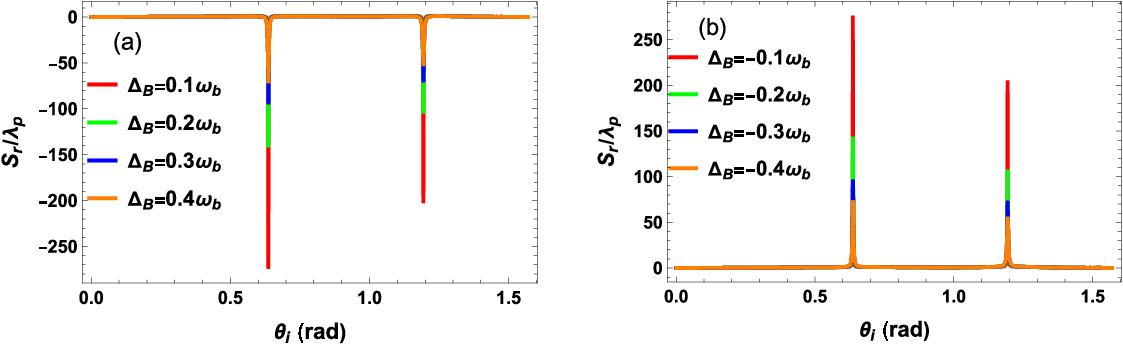}
\caption{Variation of the normalized GHS $S_r/\lambda_p$ with the probe-field incidence angle $\theta_i$ at resonance ($x=0$). Panel (a) corresponds to positive Barnett-induced magnon frequency shifts: $\Delta_B=0.1\omega_b$ (red), $0.2\omega_b$ (green), $0.3\omega_b$ (blue), and $0.4\omega_b$ (orange). Panel (b) shows the corresponding negative shifts: $\Delta_B=-0.1\omega_b$ (red), $-0.2\omega_b$ (green), $-0.3\omega_b$ (blue), and $-0.4\omega_b$ (orange). All other parameters are the same as those used in Fig.~\ref{fig3}.}
\label{fig5}
\end{figure*}

From Eqs.~(\ref{phase}) and~(\ref{GHS}), the GHS $S_r$ is determined by the complex reflection coefficient $R(k_z,\omega_p)$, particularly through its phase $\phi_r$. To investigate the effect of the Barnett frequency shift $\Delta_B$ on the GHS, we plot the magnitude of the reflection coefficient $|R(k_z,\omega_p)|$, the reflection phase $\phi_r$, its angular derivative $d\phi_r/d\theta_i$, and the normalized GHS $S_r/\lambda_p$ as functions of the probe-field incident angle $\theta_i$ in Figs.~\ref{fig3}(a)--(d), respectively. Figure~\ref{fig3}(a) shows $|R(k_{z},\omega_{p})|$ for three cases: zero Barnett shift ($\Delta_{B}=0$) and positive and negative Barnett shifts ($\Delta_{B}=\pm0.2\omega_{b}$). In all three cases, the reflection dips occur at nearly the same incident angles. Figure~\ref{fig3}(d) shows the corresponding normalized GHS, which exhibits pronounced peaks at these reflection features. For $\Delta_{B}=0$, the GHS remains very small, as shown in the inset of Fig.~\ref{fig3}(d) (red curve), whereas substantially larger GHS values are obtained for $\Delta_{B}=\pm0.2\omega_{b}$. The suppression at $\Delta_{B}=0$ is caused by the magnon--phonon coupling $G_{mb}$, which enhances absorption at resonance ($x=0$) and thereby reduces the phase gradient of the reflection coefficient~[Fig.~\ref{fig3}(c)]~\cite{waseem_Goos_2024}. When the Barnett shift is activated ($\Delta_{B}=\pm0.2\omega_{b}$), the phase slope of $R(k_{z},\omega_{p})$ at the reflection feature becomes significantly steeper~[Fig.~\ref{fig3}(b)], producing a marked increase in $S_{r}/\lambda_p$. The sign of the GHS reverses between $\Delta_{B}=+0.2\omega_{b}$ and $\Delta_{B}=-0.2\omega_{b}$, because the Barnett shift displaces the dispersive phase feature to opposite sides of the resonance, yielding opposite phase gradients~[Fig.~\ref{fig3}(c)]. Furthermore, the GHS magnitude is not symmetric for the two rotation directions: $\Delta_{B}=-0.2\omega_{b}$ produces a comparatively larger $|S_{r}/\lambda_p|$ than $\Delta_{B}=0.2\omega_{b}$, such that $|S_{r}(-\Delta_{B})/\lambda_p| \neq |S_{r}(\Delta_{B})/\lambda_p|$. This asymmetry is further quantified and discussed using the contrast ratio below.
\begin{figure}
\centering
\includegraphics[width=0.95\linewidth]{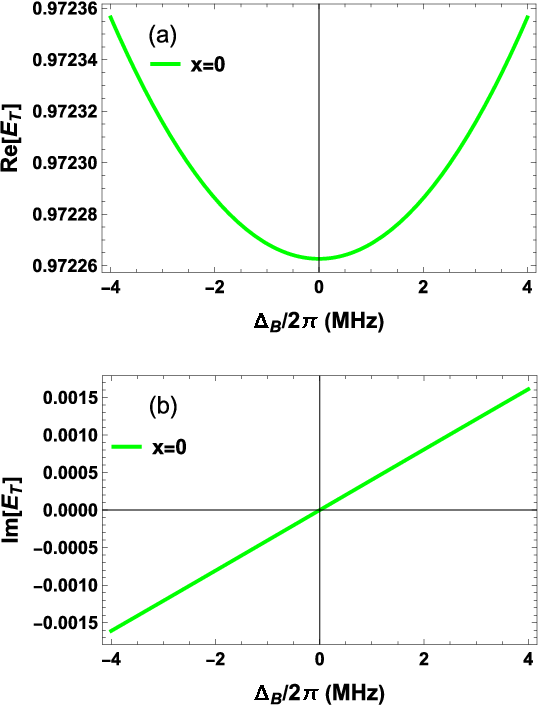}
\caption{(a) $\mathrm{Re}[E_T]$ (absorption) and (b) $\mathrm{Im}[E_T]$ (dispersion) spectra versus the Barnett shift $\Delta_B/2\pi$ (MHz) at the resonance condition $x=0$. Other parameters follow those in Fig.~\ref{fig3}.}
\label{fig6}
\end{figure}

To quantitatively assess the difference between the GHS magnitudes for opposite signs of the Barnett-induced frequency shift, we introduce a bidirectional contrast ratio $C$, defined as
\begin{equation}C(\theta_i,\Delta_B)=
\frac{\left|\left|S_r(\theta_i,-\Delta_B)/\lambda_p\right|-\left|S_r(\theta_i,\Delta_B)/\lambda_p\right|\right|}{\left|S_r(\theta_i,-\Delta_B)/\lambda_p\right|+\left|S_r(\theta_i,\Delta_B)/\lambda_p\right|}.
\end{equation}
Here, $C(\theta_i,\Delta _B)=C=1$ corresponds to the ideal case of maximum magnitude asymmetry, in which the GHS magnitude vanishes for one sign of $\Delta_B$ while remaining finite for the opposite sign. In contrast, $C=0$ indicates equal GHS magnitudes for the two opposite Barnett shifts. We evaluate $C$ over the same parameter region where the sign-dependent GHS response is observed, as shown in Fig.~\ref{fig4}. The generally small contrast ratio over most of the parameter range indicates that the GHS magnitudes for positive and negative Barnett shifts are nearly equal, with negligible differences. However, narrow regions of enhanced contrast appear near the characteristic incident angles, i.e., $\theta_i\approx0.634~\mathrm{rad}$ and $ 1.198~\mathrm{rad}$, where the magnitude asymmetry between the two rotation directions becomes more pronounced.

Next, we examine the effect of increasing the magnitude of the Barnett shift on the GHS for both positive and negative shifts, as shown in Figs.~\ref{fig5}(a) and (b). As the Barnett shift increases from $\Delta_B=0.1\omega_b$ to $0.4\omega_b$, the GHS magnitude at resonance ($x=0$) decreases monotonically. The same trend holds for negative shifts: as $\Delta_B$ changes from $-0.1\omega_b$ to $-0.4\omega_b$, the magnitude of the GHS also decreases. Physically, this behavior arises because increasing $|\Delta_B|$ shifts the magnon-induced dispersive feature away from the resonance condition, as illustrated in Fig.~\ref{fig6}. The resulting increase in probe-field absorption is accompanied by a reduced phase gradient at the resonance angle, leading to a smaller GHS magnitude. This reduction is associated with the decreased mode overlap between the coupled system and the probe field.
\begin{figure*}
\centering
\includegraphics[width=0.98\linewidth]{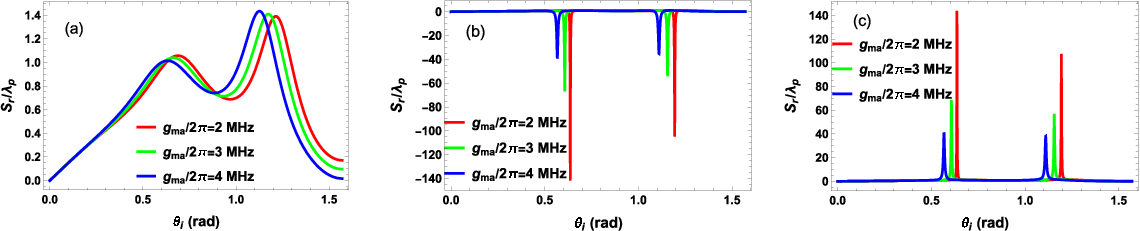}
\caption{Dependence of the normalized GHS $S_r/\lambda_p$ on the probe-field incidence angle $\theta_i$ for (a) $\Delta_B=0$, (b) $\Delta_B=0.2\omega_b$, and (c) $\Delta_B=-0.2\omega_b$ at resonance ($x=0$), considering three different photon-magnon coupling strengths, $g_{ma}/2\pi=2$, $3$, and $4~\mathrm{MHz}$. Other parameters are identical to those used in Fig.~\ref{fig3}.}
\label{fig7}
\end{figure*}
\begin{figure*}
\centering
\includegraphics[width=0.98\linewidth]{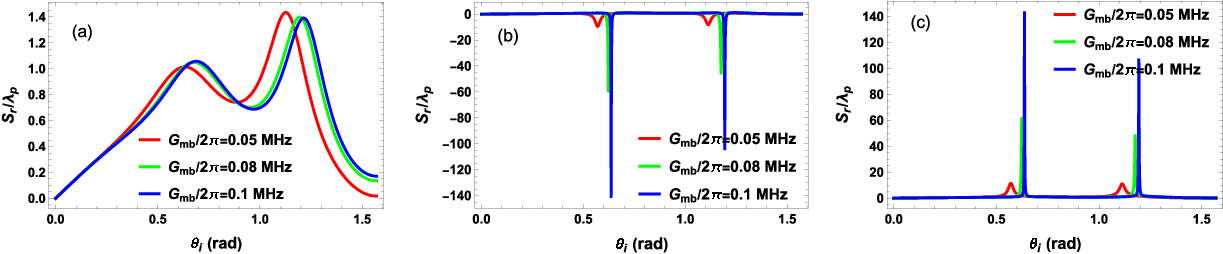}
\caption{Normalized GHS $S_r/\lambda_p$ versus the probe incidence angle $\theta_i$ for (a) $\Delta_B=0$, (b) $\Delta_B=0.2\omega_b$, and (c) $\Delta_B=-0.2\omega_b$ at resonance ($x=0$), considering three different magnomechanical coupling strengths, $G_{mb}/2\pi=0.05$, $0.08$, and $0.1~\mathrm{MHz}$. The rest of the parameters are maintained exactly as specified in Fig.~\ref{fig3}.}
\label{fig8}
\end{figure*}
\begin{figure*}
\centering
\includegraphics[width=0.98\linewidth]{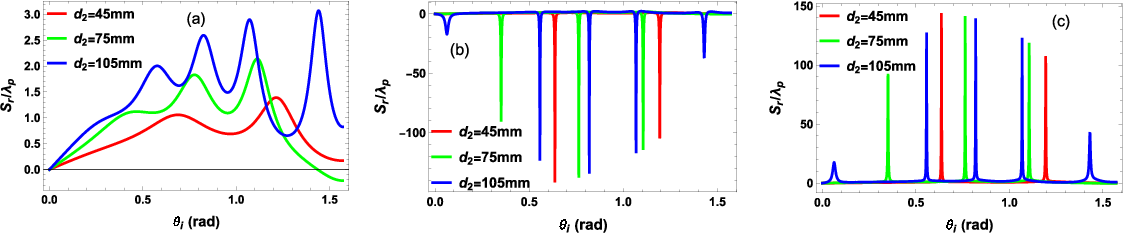}
\caption{Normalized GHS $S_{r}/\lambda$ versus the incident angle $\theta_{i}$ at resonance ($x=0$) for (a) $\Delta_B=0$, (b) $\Delta_B=0.2\omega_b$, and (c) $\Delta_B=-0.2\omega_b$. The red, green, and blue curves represent the intracavity lengths $d_2 = 45~\mathrm{mm}$, $75~\mathrm{mm}$, and $105~\mathrm{mm}$, respectively. Other parameters remain unchanged from Fig.~\ref{fig3}.}
\label{fig9}
\end{figure*}

Based on the preceding analysis of the sign-dependent GHS response induced by the Barnett frequency shift, we next examine how the coupling parameters of the CMM system affect the normalized GHS. In the considered CMM system, two types of interactions are present: photon--magnon coupling ($g_{ma}$) and magnomechanical coupling ($G_{mb}=g_{mb}m_s$). First, we investigate the influence of $g_{ma}$ on the normalized GHS $S_r/\lambda_p$ in the absence and presence of the Barnett-induced frequency shift $\Delta_B$ at a constant $G_{mb}/2\pi=0.1~\mathrm{MHz}$, as shown in Figs.~\ref{fig7}(a--c). For $\Delta_B=0$, increasing $g_{ma}/2\pi$ from $2$ to $4~\mathrm{MHz}$ progressively modifies the angular dependence of the GHS, particularly the position and magnitude of its peaks, owing to enhanced photon--magnon hybridization [Fig.~\ref{fig7}(a)]. Physically, stronger photon--magnon coupling facilitates more efficient energy exchange between the two subsystems, thereby modifying the phase response of the reflected probe field and, consequently, the magnitude of the GHS. In the presence of the Barnett shift, $\Delta_B=\pm0.2\omega_b$, the GHS magnitude is gradually reduced with increasing $g_{ma}$. This suppression can be attributed to the Barnett-induced modification of the magnon resonance frequency, which alters the photon--magnon hybridized resonance condition and weakens the angular variation of the reflection phase. Consequently, the phase gradient $\partial\phi_r/\partial\theta_i$ is reduced, leading to a smaller GHS magnitude~[Figs.~\ref{fig7}(b) and (c)].

Second, we investigate the effect of $G_{mb}$ on $S_r/\lambda_p$ in the presence and absence of $\Delta_B$, for a fixed $g_{ma}/2\pi=2~\mathrm{MHz}$, as shown in Figs.~\ref{fig8}(a)--(c). At $\Delta_B=0$, increasing $G_{mb}/2\pi$ from $0.05$ to $0.1~\mathrm{MHz}$ leads to a reduction in the GHS magnitude [Fig.~\ref{fig8}(a)]. This behavior is associated with increased probe-field absorption at resonance ($x=0$), which modifies the complex reflection coefficient~[Eq.~(\ref{Reflection-coefficent})] and, consequently, reduces the phase gradient responsible for the GHS, consistent with the findings reported in Ref.~\cite{waseem_Goos_2024}. In contrast, for $\Delta_B=\pm0.2\omega_b$, the GHS magnitude increases with $G_{mb}$. This enhancement can be attributed to the stronger magnomechanical interaction, which, together with the Barnett-induced magnon-frequency shift, modifies the magnon--phonon hybridization and enhances the angular dispersion of the reflection phase, thereby increasing the GHS magnitude. Moreover, the direction of the GHS is reversed for the two opposite signs of $\Delta_B$ [Figs.~\ref{fig8}(b) and (c)].

Finally, the GHS exhibits a strong dependence on the geometric configuration of the cavity, particularly on the total cavity thickness $L=2d_1+d_2$. Consequently, precise control of the cavity dimensions is essential for reliable manipulation of the beam shift. Figures~\ref{fig9}(a)--(c) show the angular dependence of the normalized GHS $S_r/\lambda_p$ at resonance ($x=0$) for three different intracavity lengths, $d_2=45~\mathrm{mm}$, $75~\mathrm{mm}$, and $105~\mathrm{mm}$, considering both the absence and presence of the Barnett-induced frequency shift $\Delta_B$. For $\Delta_B=0$, increasing $d_2$ leads to a larger number of GHS peaks and an overall enhancement of their magnitudes. The emergence of additional peaks arises from the increased propagation phase accumulated by the probe field in the longer intracavity region, resulting in more pronounced angular variations of the reflection phase~[Fig.~\ref{fig9}(a)]. In contrast, for $\Delta_B=\pm0.2\omega_b$, increasing $d_2$ also produces additional GHS peaks, but with reduced amplitudes. This behavior stems from the combined effects of the increased propagation phase and the Barnett-induced modification of the magnon resonance frequency, which alters the interference between the cavity and magnon-mediated pathways. Consequently, the angular phase variation becomes less pronounced near the individual resonance features, leading to a reduced GHS magnitude for both positive and negative values of $\Delta_B$~[Figs.~\ref{fig9}(b) and (c)].

\section{Conclusion}
We have proposed a scheme for realizing a tunable nonreciprocal GHS in a CMM system through the Barnett effect. In the absence of the Barnett-induced frequency shift, the probe field exhibits magnomechanically induced transparency (MMIT) near resonance. A finite Barnett shift displaces the transparency window from resonance, with the direction of the displacement determined by the sign of the shift. For a fixed rotation direction, this sign can be reversed by switching the direction of the bias magnetic field.

We have further analyzed the GHS of the reflected probe field. When the Barnett shift vanishes, the GHS remains relatively small because the probe field experiences strong resonant absorption~\cite{waseem_Goos_2024}. Activating the Barnett shift substantially enhances the GHS, while reversing its sign reverses the direction of the spatial displacement. The GHS does not, however, increase monotonically with the magnitude of the Barnett shift. At sufficiently large shifts, the system moves away from the spectral region of strong phase dispersion, reducing the angular derivative of the reflection phase and, consequently, the GHS. We have also shown that the magnon--photon and magnon--phonon couplings affect the GHS in opposite ways, with their respective roles depending on whether the Barnett shift is present. In addition, the cavity length provides an independent parameter for controlling the magnitude of the nonreciprocal GHS.

The nonreciprocal response originates from the sign-dependent modification of the effective magnon frequency by the Barnett effect. Opposite bias-field directions therefore produce different dispersive phase responses for a fixed rotation direction, leading to distinct GHSs of the reflected probe field. This mechanism enables magnetic control of both the magnitude and direction of the beam displacement without requiring structural asymmetry. Our results thus establish a theoretical framework for Barnett-mediated nonreciprocal beam manipulation in CMM systems, with potential applications in reconfigurable microwave photonic devices and Barnett-field sensing.

From an experimental perspective, CMM systems constitute a well-established platform for investigating hybrid magnonic, mechanical, and microwave phenomena. Their demonstrated applications include microwave-to-optical transduction~\cite{Shen2022}, the generation of magnonic frequency combs~\cite{Xu2023}, magnetometry~\cite{Colombano2020}, and mechanical bistability~\cite{Shen2022MB}. Recent experiments have also shown that an ultrafast Barnett effect generated by circularly polarized phonons can induce remote magnetization switching~\cite{Davies2024Phononic}. These advances provide a promising foundation for implementing the proposed Barnett-mediated control of the GHS.

Controlled rotation of the YIG sphere could be achieved by attaching it to a thin rod driven by a mechanical rotor or turbine, following techniques developed for rotating optical resonators~\cite{Maayani2018}. High rotation frequencies have also been demonstrated using levitated spherical and nanoscale rotors~\cite{Marcel2018,Reimann2018}. Alternatively, an air-turbine drive could reduce electromagnetic interference from the rotation mechanism~\cite{Ono2015}. The magnitude of the Barnett shift can be tuned through the rotation speed, whereas its sign can be reversed by changing either the rotation direction or the orientation of the bias magnetic field. Experimentally observed rotation-frequency fluctuations of approximately $\pm 30~\mathrm{Hz}$ at a rotation frequency of $1.6~\mathrm{kHz}$~\cite{Ono2015} indicate that sufficiently stable rotation is achievable. Although incorporating a rotating YIG sphere may require modifications to existing CMM architectures, these established rotation techniques provide a viable route toward the experimental realization of the proposed nonreciprocal GHS.

\section*{Acknowledgement}
We acknowledge the financial support from the NSFC under Grant No. 12174346.\\

\section*{references}

%

\end{document}